\documentstyle[aps,prl,preprint,tighten]{revtex}
\input psfig
\begin{document}
%\preprint{Original version by Peter}
\draft
\title{\bf
\vspace*{-10mm}
\begin{flushright}
{\large Applied Physics Report 95-14}
\end{flushright}
%Universal
Excess Noise in Biased Superconducting Weak Links}
\author{J.~P. Hessling\thanks{Electronic address:
hessling@fy.chalmers.se} $^{(1)}$ \and V.~S. Shumeiko $^{(1)}$
\and Yu.~M. Galperin $^{(2)}$
\and G. Wendin $^{(1)}$}
\address{$^{(1)}$
Department of Applied Physics, Chalmers University of
Technology and G\"{o}teborg University, \\
S-412 96, G{\"o}teborg, Sweden}
\address{$^{(2)}$
Department of Physics, University of Oslo,
P.O. Box 1048 Blindern, N 0316 Oslo 3, Norway, \\
and A.F. Ioffe Physico-Technical Institute, 194021 St. Petersburg,
Russia\\
\mbox{\rm \today}
}
\maketitle

\begin{abstract}
Non-equilibrium excess noise of a short quasi one-dimensional
constriction between two superconductors is considered.
A general expression for the current-current correlation function
valid for arbitrary temperatures and bias voltages is derived.
This formalism is applied to a current-carrying quantum channel with
perfect transparency.
Contrary to a transparent channel separating two normal conductors,
a weak link between two superconductors exhibits a finite level of noise.
The source of noise is fractional
Andreev scattering of quasiparticles with energies $|E|$ greater
than the half-width $\Delta$ of the superconducting gap.
For high bias voltages, $V \gg \Delta /e$, the
relation between the zero-frequency limit of the noise
spectrum, $S(0)$, and  the excess current $I_{\text{exc}}$ reads
$S(0)=(1/5)|e|I_{\text{exc}}$. As $\Delta \rightarrow 0$ both the excess
noise and the excess current vanish linearly in $\Delta$,
%$\propto \Delta$,
their ratio being constant.
%\parbox{14cm}{\medskip \rm \indent
\end{abstract}


%\section{Introduction}
%\section*{}
\paragraph*{Introduction.}
%\hspace*{5mm}
As is well known ((see, e.g., Ref.~\onlinecite{b1}), electron transport
through short superconducting weak links  can be described in terms of
multiple Andreev reflections (MAR). Calculations of dc $I-V$ curves of
voltage-biased weak links \cite{b2,b3,b4,b5,b6,b7}
have predicted specific  current singularities at voltages $V=
2\Delta /en, \; n=1,2,...,$ where $2\Delta$ is the superconducting
energy gap and $e$ the electronic charge. This feature is a
manifestation of MAR. However, dc properties
carry only indirect information about weak link dynamics --
more direct information can be found from theoretical studies of ac currents.
Studies of ac transport
have mostly been carried out for large voltages, $eV \gg \Delta$,
the limitation being caused by the fact that at low voltages it is
necessary to take into account a large number of MAR.
Recently, Averin and Bardas\cite{ab} have considered a
model of a short constriction between two superconductors which
permits quantitative calculations of ac currents for arbitrary voltages
$V$ and transparencies $D$.  Unfortunately, experimental studies of ac
characteristics are difficult.

In addition to studies of transport properties,
investigations of non-equilibrium excess noise
provide qualitative information about the dynamics of mesoscopic
junctions.  This type of noise can be investigated without ac bias,
and directly probes the correlation
between particles passing through the junction. It is not surprising
that noise properties of weak links of the Josephson type
have been extensively discussed (see, e.g., Refs~\onlinecite{rs,likh1}
for a review).  The specific role of quantum mechanical correlations in
normal contacts due to the Fermi statistics of electrons was
considered in Refs.~\onlinecite{lb1,lb2,lb3}.  There it was
found a specific quantum reduction of noise in
comparison with the so-called Schottky limit \cite{schottky}.

In the present study we discuss noise properties of a short
voltage-biased constriction separating two superconductors.
The junction is modeled as an
adiabatically-smooth point contact with typical length $L$ much
larger than the Fermi wave length $k_{\text F}^{-1}$, but much less than
the coherence length of the superconductors. In this limit
when $k_{\text F}L \gg 1$, one can neglect intermode scattering within
the constriction and
write the electron wave function as a superposition of different
independent modes for the transverse motion.
As a result, both current and noise is a sum of independent
contributions from each mode. Each term corresponding to a certain mode
can be expressed through
the solution of the scattering problem for electronic states in the leads.
In the following,
a general theoretical approach to non-equilibrium noise in
voltage-biased contacts is presented and noise is given
in terms of the solution of the scattering problem.

The case of perfect transparency, $D=1$,
is particularly interesting because then there is no
excess noise in the normal state (see below).
In this limit we give an explicit result for large bias voltages,
$eV \gg \Delta$.
In the superconducting state there is a {\em finite} level of excess
noise due to Andreev reflection. The physical origin is
that the amplitude for this type of reflection %in a biased junction
is non-zero but less than one,
for quasiparticle states outside the gap, i.e.
having energies $|E| > \Delta
%, \ |E-\Delta|\ll \Delta
$.
The quasiparticle then have a finite probability to be
reflected from, {\em or} transmitted through, the junction which is required
in order to have a fluctuating current.
We show that the zero-frequency limit of the noise spectrum is proportional to
the excess current\cite{b1,b2}, the
proportionality coefficient being $e/5$. %{\em universal}.

This letter is organized as follows. First, a formulation of the general
scattering problem will be given and the solution for $D=1$
specified. Then the excess noise will be
expressed in terms of the scattering solution,
and finally we apply this to the limiting case $D=1$, zero temperature and
zero frequency in the large voltage limit.

%\section{Scattering theory of voltage-biased weak links}
%\label{scatt}
\paragraph*{Scattering problem.}% \hspace*{5mm}
In the original Landauer picture\cite{land}, scattering in a normal
voltage-biased  point contact is elastic. The voltage $V$ only
enters as a relative displacement of the Fermi levels
in the reservoirs. For a contact between superconductors, an
alternative picture where the voltage $V$ is included in the phase
of the wave function is more convenient. According to this picture,
constant electric potentials in the leads far from the contact are
removed by local gauge transformations which are different on
the two sides of the junction. Consequently, the voltage is
included as phase factors in all scattering states.
Transmission through the contact now formally appears as inelastic and
for a normal contact there is one transmitted state at
energy $E+eV$  with amplitude $d$, and one reflected state with amplitude $r$
at the same energy E as the incoming  state.
% This interpretation is particularly useful for superconducting junctions.
In a voltage-biased superconducting contact though, multiple
Andreev scattering processes to all energies $E+neV$ (where  $n$ is an
integer) are possible\cite{b1}.

In the leads, the electronic states can be described as solutions to
the Bogoliubov-de Gennes (BdG) equation\cite{dg} and have the form of
two-component spinors $\Psi$ which can be specified by the side of the
junction ($L$ or $R$) as a subscript, and by the direction of the
longitudinal momentum ($\pm$) as a superscript.
Introducing slowly varying spinors $\Phi^{\pm}_{L(R)}$
the total wave function for each transverse mode will be expanded as
$\Psi= e^{- i Et/\hbar}
\sum_{\pm} e^{\pm i k_Fx} \Phi^\pm$.
The spinors $\Phi$ satisfy a  reduced time-independent BdG
equation\cite{dg}, $(\pm \hbar \tilde{k}_F \hat{p}/m_e
+ \sigma_x \Delta) \Phi^\pm = E \Phi^\pm$
where $m_e$ is the electronic mass and $\hat{p}$ the momentum operator.
The Fermi wave vector (energy) is $k_F$
$(E_F)$ outside the channel, and
$\tilde{k}_F =  \hbar^{-1}\sqrt{2 m_e E_F[1 - (n\pi/k_F w)^2]}$
for mode $n$ inside the contact of width $w$. The spinors $\Phi$
change on the scale of the  coherence length $\xi_0 =\hbar^2 k_F/\pi
\Delta m_e$, i.e. they can be considered as being constant
in the contact region.
In the following we adopt the Andreev
approximation\cite{aa} which is valid up to lowest order in the
small parameter $\Delta/E_{\text F}$. In this approximation
the scattering is described by the boundary condition\cite{b5,ab},
\begin{equation}\label{scattform}
\left(
\begin{array}{c}
\Phi_L^{-} \\
\Phi_R^{+}
\end{array}
\right)
=
\hat{V}
\left(
\begin{array}{c}
\Phi_L^{+} \\
\Phi_R^{-}
\end{array}
\right) \, , \,
\hat{V}=
\left(
\begin{array}{cc}
\hat{r} & \hat{d}e^{-i \hat{\sigma}_z eVt/\hbar} \\
\hat{d}e^{i \hat{\sigma}_z eVt/\hbar} & {-\hat{r}^\star d/d^\star}
\end{array}
\right)\, .
\end{equation}
The parameter $r(d)$ is the normal reflection (transmission)
amplitude, while the matrix $\hat{r} (\hat{d})$ is the same quantity
multiplied by the $2 \times 2$ unit matrix.
Since the junction length is much less than the
coherence length\cite{b10},  the coordinate
dependence of the phase of the order  parameter can be approximated by
a step function  centered at the junction, and the magnitude of the
order parameter  by a constant.
This allows us to include the  phase difference $eVt/\hbar$
in the matching matrix $\hat{V}$, see above.
The general boundary condition (\ref{scattform}) has to be applied to the
four distinct scattering cases -- electron- and hole-like quasi particles
incident from the left and the right, and for positive as well as for
negative energies.

\begin{figure}
\centerline{\psfig{figure=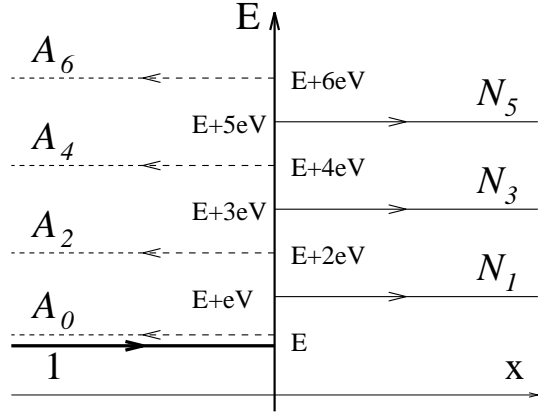,width=7cm}}
\vspace*{10mm}
\caption{ \protect{\label{fig:d1state}}
A scattering state for the perfect transmission $(d=1)$ case.
Full lines represents one quasiparticle
character $e (h)$ while dashed
lines means the opposite character $h (e)$
-- full thick line corresponds to incoming quasiparticle.
Some Andreev reflection ${\cal A}_n$ and
normal transmission amplitudes ${\cal N}_n$ to sidebands $n$
with energy $E+neV$ are shown.
Interchanging $e$ and $h$ characters, or
changing the sign of the applied
bias voltage,
or interchanging left and right,
causes the scattering state to be mirror-imaged
around the energy $E$
of the incoming particle, $n \rightarrow -n$.}
\end{figure}

The scattering problem substantially simplifies for the case of perfect
transmission, $d=1$.
The wave vector
is then conserved to lowest order in $\Delta/E_F$.
Hence only wave vector conserving scattering events can take place,
that is, Andreev reflection to energies $E+2neV$ with
amplitudes ${\cal A}_{2n}$ and
normal transmission to energies $E+(2n+1)eV$ with
amplitudes ${\cal N}_{2n+1}$.
Further, the direction of energy transfer is prescribed by the sign
of the voltage and by the scattering case. For a given
scattering case only scattering to higher ({\em or} lower)
energies is possible.
A typical scattering state is shown in Fig.~\protect{\ref{fig:d1state}}.
The scattering amplitudes ${\cal A}_{2n}$ and ${\cal N}_{2n+1}$
 can be expressed as
\begin{equation}
{\cal A}_{2n}= X_{2n}\, , \quad
{\cal N}_{2n+1}= X_{2n+1}\, ,
\end {equation}
through the function $X_n$,
\begin{equation}
X_{n} \equiv \left\{
\begin{array}{lcl}
[ (u_0^2 - v_0^2)/{v_0 u_n}]\prod_{k=0}^{n-1} ({v_k}/{u_k}),
&& n \neq 0 \\
- {v_0}/{u_0},  && n = 0 \, .
\end{array}
\right.
\end{equation}
with upward (downward) recursion, $n \geq 0$, for scattering states associated
with incident electron-like (hole-like) quasiparticles from
the left and hole-like (electron-like) ones from the right.
The coherence factors $u_n\equiv u(E+neV), v_n\equiv v(E+neV)$ are
for $E>0$ given by
$u^2(E)= \min{[1,E/\Delta]} (1+\sqrt{1-\Delta^2/E^2})/2$ and
$v^2(E)= \min{[1,E/\Delta]}(1-\sqrt{1-\Delta^2/E^2})/2$;
for $E<0$, $u(E)= -v^\star (-E)$ and $v(E)=u^\star (-E)$.
In the large voltage limit $eV \gg \Delta$ only
 three scattering amplitudes have to be taken into account,
\begin{eqnarray}\label{ampl}
{\cal A}_0  & = & -v_0/u_0, \quad
{\cal A}_{\pm 2}  =  -v_{\pm 1}/u_{\pm 1}, \quad
{\cal N}_{\pm 1}  =  (u_0^2 -v_0^2)/(u_0 u_{\pm 1}) \, .
\end{eqnarray}
All other amplitudes are at most of order $\Delta/eV$ for all energies.

\paragraph*{Excess Noise.} %\hspace*{5mm}
Current noise is usually defined as the Fourier transform
of the current-current correlation function, $S(\tau)=1/2 \langle
\left[ \hat{I}(t+\tau), \hat{I}(t) \right]_+ \rangle -
\langle \hat{I}(t)\rangle^2 $,
where $[A,B]_+ = AB + BA$.
Using the Bogoliubov-de Gennes transformation\cite{dg},
we can express electron field operators in terms of
quasiparticle operators $\gamma$. The current operator will
be
\begin{eqnarray}\label{curroper}
\hat{I}(t)&=& \frac{\hbar e}{2m_e} \sum_{{\bar 1},{\bar 2}}
\left[
{\hat \gamma}_{{\bar 1}}^{\dagger} {\hat \gamma}_{{\bar 2}}
j_+({\bar 1},{\bar 2},t)+
{\hat \gamma}_{{\bar 1}} {\hat \gamma}_{{\bar 2}}^{\dagger}
j_-({\bar 1},{\bar 2},t)
\right.
%\nonumber \\ &&
+\left.
{\hat \gamma}_{{\bar 1}} {\hat \gamma}_{{\bar 2}}
\sigma_{\bar 1} j_x({\bar 1},{\bar 2},t) +
{\hat \gamma}_{{\bar 1}}^{\dagger} {\hat \gamma}_{{\bar 2}}^{\dagger}
\sigma_{\bar 2}
j_x^{\star}({\bar 2},{\bar 1},t)
\right]\, .
\end{eqnarray}
The sum is over quantum labels ${\bar 1}$ and ${\bar 2}$
ranging over {\em all} states
(meaning scattering case index, energy of incoming state,
side bands $n$, and spins $\sigma$).
Above we have defined three different types of generalized currents,
\begin{eqnarray}\label{gencurr1}
j_+({\bar 1},{\bar 2},t) & \equiv & \langle \Phi({\bar 1}) |
i \nabla \hat{\sigma}_{11} \Phi({\bar 2}) \rangle +
\langle i \nabla \Phi({\bar 1}) |
\hat{\sigma}_{11} \Phi({\bar 2}) \rangle, \\
j_-({\bar 1},{\bar 2},t) & \equiv & \langle \Phi^\star({\bar 1}) |
i \nabla \hat{\sigma}_{22} \Phi^\star({\bar 2}) \rangle +
\langle i \nabla \Phi^\star({\bar 1}) |
\hat{\sigma}_{22} \Phi^\star({\bar 2}) \rangle, \\
%\langle i \nabla \Phi({\bar 2}) |
%\hat{\sigma}_{22} \Phi({\bar 1}) \rangle -
%\langle \Phi({\bar 2}) |
%i \nabla \hat{\sigma}_{22} \Phi({\bar 1}) \rangle, \\
j_x({\bar 1},{\bar 2},t) & \equiv & \langle \Phi^\star({\bar 1}) |
i \nabla \hat{\sigma}_{21} \Phi({\bar 2}) \rangle +
\langle i \nabla \Phi^\star({\bar 1}) |
\hat{\sigma}_{21} \Phi({\bar 2}) \rangle \, , \label{gencurr3}
\end{eqnarray}
expressed through the $2 \times 2$
matrices $\hat{\sigma}_{ij}$
having element $(i,j)$ equal 1
and zero otherwise.
The scalar product between two spinors is here defined as
%(assume $\Phi= \left( \begin{array}{c} u \\  v \end{array} \right)$),
(assume $\Phi^T= (u, v)$),
$%\begin{equation}
\langle \Phi ({\bar 1}) |
\Phi ({\bar 2}) \rangle \equiv u^\star({\bar 1}) u({\bar 2}) +
v^\star({\bar 1}) v({\bar 2})
$. %\end{equation}
Introducing the normalized BCS density of states per spin
for incoming quasiparticles,
$\tilde{g}(E) =[|E|/\sqrt{E^2 - \Delta^2}]
\Theta(|E| - \Delta)$,
our general expression for noise can be written as\cite{details}
\begin{eqnarray}\label{gennoise}
S(\omega,V,T) & = & \frac{e^2}{8 \pi^2 \hbar}\sum_{\pm , n}
\int_{-\mu}^{\infty}
f(E)[1-f(E + neV \pm \hbar \omega)]
\nonumber \\ && \times \,
\tilde {g}(E) \tilde {g}(E + neV \pm
\hbar \omega)
{\text {Tr}}\, \hat{\Omega}^{(n)}(E,E + neV \pm \hbar \omega) \,dE \, ,
\\
\hat{\Omega}^{(n)}(E,E^{\prime}) & \equiv &
%\sum_{n} \left\{
\hat{Q}^{(n)}(E,E^{\prime})
\hat{P}(E^{\prime},E) +
%\nonumber \\ && \sum_n
\sum _{\nu=1,2} \hat{Q}^{(n)}_{\nu}(E,E^{\prime})
\hat{P}_{\nu}(E^{\prime},E) %\right\}
\, , \\
\hat{Q}^{(n)}_{(\nu)}(E,E^{\prime})
& \equiv &  \sum_{m}
\langle {\bf \Phi}_{n+m}(E) \hat{s}|\bigotimes (\hat{\sigma}_{\nu \nu})
{\bf \Phi}_m (E^{\prime}) \rangle, \\
\hat{P}_{(\nu)}(E^{\prime},E)
& \equiv &
\langle \sum_{n} {\bf \Phi}_n (E^{\prime}) \hat{s}|\bigotimes
(\hat{\sigma}_{\nu \nu}) \sum_{m} {\bf \Phi}_m (E) \rangle \, .
\end{eqnarray}
Above, ${\bf \Phi}_m(E)$ is a row vector
for harmonic side-band $m$
consisting of four two-component (column)
spinors $\Phi_m(E)$,
each corresponding
to one of the four scattering cases.
The symbol $\bigotimes$ denotes outer product in
scattering case indices.
Thus $Q$ and $P$ are $4 \times 4$
matrices with element
$(i,j)$ corresponding to scattering cases $i$ and $j$ in the bra and ket
states in the definitions above.
The trace is then calculated in this four-dimensional space.
The parenthesis around $\hat{\sigma}_{\nu \nu}$
implies that $\hat{\sigma}_{\nu \nu}$ should
be excluded when calculating
$\hat{Q}^{(n)}$ and $\hat{P}$.
The operator $\hat{s}$ causes the inner product
to be multiplied by the average of the wave
vectors  (in units of the Fermi vector) % ($k=\pm k_F$ to lowest order)
of the bra and ket states --
only if the wave vectors have equal sign will the term
contribute $(\hat{s} \Rightarrow \pm 1)$.

For perfect transmission $(d=1)$, the wave vector
is conserved when quasiparticles are scattered in the junction
implying that all subbands have the
same wave vector (to leading order in $\Delta/E_F$).
This allows us to assign a value
of the sign of the wave vector ($\pm k_F$)
to each scattering state containing all its subband components.
The number of contributing terms in the trace % over scattering cases
in the noise expression is then reduced from 16 to 8
(see operator $\hat{s}$  in $S(\omega)$).
In the limit of large bias voltages $V$ all
scattering amplitudes, to lowest order in $\Delta/eV$,
are given by Eq.~(\ref{ampl}).

Putting in proper coherence factors into the expressions for
scattering amplitudes and spinors, we find that for
zero temperature,
the zero frequency limit of the noise %spectrum
can be written as
\begin{equation} \label{sexc}
S_{\text {exc}}(0) \rightarrow \frac{2e^2}{\pi^2 \hbar}
\int_{-\infty}^{-\Delta}
D_A(E) [ 1- D_A(E) ]
\, dE = \frac{8e^2 \Delta}{15\pi^2 \hbar}\, ,
\end{equation}
as $\Delta/eV \rightarrow 0$.
The quantity
$D_A(E) \equiv 1 - {\cal A}^2_0$
can be interpreted as the elastic Andreev
transparency at energy $E$. As a function of $E$, $D_A=0$ for
$|E| \leq \Delta$
(complete Andreev reflection). It is close to 1 for  $E \gg
\Delta$, while in the vicinity of the gap for
$|E| > \Delta $,
both  $D_A$ and $1- D_A$ are finite.
For arbitrary voltages, noise also contains inelastic
contributions. In the limit of large voltages though, these terms vanish
because of the decay of the Andreev reflection probability ${\cal A}^2_0$
far away from the gap region $|E| \leq \Delta$.
This voltage independent level of noise is proportional
to the excess current\cite{b1,b2},
$S_{\text {exc}}(0)= |e I_{\text {exc}}|/5 $.
It is illuminating to compare this with
the level of noise in a
normal quantum point
contact\cite{lb1,lb2,lb3} at $\omega=0$ and  $T=0$,
\begin{equation}
S_N(0) = \frac{e^2}{\pi^2 \hbar} \int_{-eV}^{0} D_N(E) (1-D_N(E)) \, dE \, ,
\end{equation}
where $D_N(E)$ is the normal transparency.
In our case, $D_N=|d|^2=1$, and $S_N$ vanishes.

\paragraph*{Conclusion.}%\hspace{5mm}
We have shown that there is a finite level of current noise
in voltage-biased junctions between %two
superconductors, even if the  transparency in the normal state is
perfect. This excess noise is induced by fractional
Andreev reflection and for large
bias voltages ($eV \gg \Delta$) it is proportional to the excess current,
the proportionality coefficient being %universal and
equal to
$|e|/5$. Consequently, the relationship between the excess noise and
the excess current is qualitatively similar to the classical relation between
shot noise and average current in the {\em low}-voltage regime,
$S=2|e|I$. The quantum reduction of
excess noise (coefficient 1/5) is in our case pronounced.

This work was supported by NorFA,
grant no. 94.35.174-O, the Nordic Research Network on the Physics of
Nanometer Electronic Devices,
the Swedish Natural Science Research Council (NFR),
and the Swedish Superconductivity Consortium.
JPH gratefully acknowledges the hospitality
of the Department of Physics at Oslo
University, Norway.

\begin{references}
\bibitem{details} Details of the derivation of noise
will be published elsewhere.
\end{references}
\end{document}